# Realization of Photonic Charge-2 Dirac Point by Engineering Super-modes in Topological Superlattices

Mengying Hu,[1] Kun Ding,[2] Tong Qiao,[1] Xi Jiang,[1] Qiang Wang,[1] Shining Zhu,[1] and Hui Liu[1,★]

[1]*National Laboratory of Solid State Microstructures, School of Physics, Collaborative Innovation Center of Advanced Microstructures, Nanjing University, Nanjing 210093, China*

[2]*The Blackett Laboratory, Department of Physics, Imperial College London, London SW7 2AZ, United Kingdom*

★ Email: liuhui@nju.edu.cn

**Abstract**

Quite recently a novel variety of unconventional fourfold linear band degeneracy points has been discovered in certain condensed-matter systems. Contrary to the standard 3-D Dirac monopoles, these quadruple points referred to as the charge-2 Dirac points are characterized by nonzero net topological charges, which can be exploited to delve into hitherto unknown realms of topological physics. Here, we report on the experimental realization of the charge-2 Dirac point by deliberately engineering hybrid topological states called super-modes in a 1-D optical superlattice system with two additional synthetic dimensions. Utilizing direct reflection and transmission measurements, we exhibit the existence of super-modes attributed to the synthetic charge-2 Dirac point, which has been achieved in the visible region for the first time. We also show the experimental approach to manipulating two spawned Weyl points that are identically charged in synthetic space. What's more, topological end modes uniquely resulting from the charge-2 Dirac point can be delicately controlled within truncated superlattice samples, opening a pathway for us to rationally engineer local fields with intense enhancement.

## Section I. Introduction

Ever since the remarkable discovery that fermion-like energy excitations predicted by relativistic quantum field theories can emerge in periodically electronic crystals whose band structures display linear band degeneracy points, a great deal of theoretical and experimental interest has been attracted in exploring such materials known as topological semimetals. The corresponding gapless semimetal phases are regarded as novel topological states, which open a new era in investigating condensed matter physics. Substantial attention is engaged by Weyl points (WPs) [1-6] and 3-D Dirac points (DPs) [7, 8]. WPs that reported actively in electronic systems are identified as synthetic magnetic monopoles in momentum space, carrying topological charges (Chern numbers) of $\pm 1$ and featured by "Fermi arc" surface states [9, 10]. DPs can be viewed as two overlapping WPs with opposite topological charges, predicted and observed in crystals as well. However, it has been recently demonstrated that unconventional topological points appear in certain crystal structures [11-19], which cannot be described in accordance with an emergent relativistic field theory. On such candidate is the charge-2 Dirac point (CDP), existing as a double-Weyl phonon in transition metal monosilicides [15, 16]. It's generated by merging a pair of identically charged WPs, and hence possessing the topological charge of $\pm 2$. Consequently, CDPs are radically distinct from traditional DPs and can give rise to novel physical phenomena.

While topological semimetals found in nature exhibit exotic phases of matter, great process in understanding such band topology has also been impelled by the research on engineered systems. The core idea of engineering lattices is to create emergent band structures analogous to those formed in electronic crystals, which can be highly tunable and have fundamentally discriminative properties, providing us unprecedented opportunities of studying topological physics. Recent developments of experimental techniques have propelled ultracold atomic gases [20, 21], photonics [22-28] and acoustics [29-32] as promising systems to engineer WPs and DPs with novel emergent properties. Constructing complex 3-D structures with certain symmetry broken is perceived as the most common strategy [22-26, 29-31], whereas another route to realize topological points is based on synthetic space [32-37]. The interest of the latter is fueled by the capacity to study topological features of these 3-D degenerate points in 2-D (1-D) systems via introducing controllable artificial dimension(s) in addition to the real spatial degree(s), and thus simplify experimental designs [20, 28, 32, 36, 38, 39]. Either of the methods has been extensively exploited for WPs and DPs in the recent years. Nevertheless, as for the CDP, the only engineered system supporting it reported so far is made up of an acoustic metamaterial corresponding to a classical 3-D phononic crystal with a non-symmorphic structure [40]. To our knowledge, engineered systems have not yet been established to realize CDPs in the optical frequency regime, neither has the concept of synthetic space. Here, we propose an experimentally acceptable scheme to realize CDPs in a 1-D optical superlattice system with working frequencies lying in the visible region harnessing synthetic dimensions, and manipulate the spawned WPs with the same topological charges. For attaining this it is crucial to design suitable photonic modes interacting with each other to form a 1-D superlattice as the first step. Instructively, interfaces between distinct topological phases of matter host robust and exotic quantum states, the use of which acts as a strong driver of current research in condensed matter [41-45]. Hence, we stack together two kinds of photonic crystals (PCs) belonging to different class of topology to create such topological interface modes (TIMs), and on this basis topological states of photons associated with CDPs can be fully investigated under the introduction of synthetic space, facilitating the experimental realization which is otherwise elusive at such frequencies.

Furthermore, intriguing topologically protected end modes emerge at the termini of the truncated superlattice, guaranteed by the CDP with nonzero topological charge. More precisely, these end modes uniquely result from the bulk-edge correspondence [46] for each of the two WPs producing the CDP in synthetic space, which in turn could be tuned independently. Such topological end modes resemble surface states in Weyl semimetals [2, 3, 6, 10], holding great potential for applications in nonlinear optics [47], quantum optics [48], and lasers [49] owing to strongly enhanced localized fields.

The rest of the paper is organized as follows: In Section II, we reveal the design concept of the realization of CDPs in our work. In Section III, we elaborate on the approaches to modulate one single TIM along with the interaction between two TIMs. In Section IV, we focus on the strategy to create the CDPs and spawned WPs in synthetic space. In Section V, we explore the topological end modes based on the established superlattice system. We conclude and discuss in Section VI.

**Section II. Design concept of the creation of CDPs**

The starting point of our scheme is to construct a 1-D topological superlattice by use of TIMs existing at interface of two PCs with discriminative topological class. Specifically, our lattice consists of these two PCs stacked alternatively, in which each interface supports a TIM that hybridizes with each other to form a novel variety of artificial collective modes, resulting in a 1-D superlattice band structure where a single TIM serves as the photonic orbital. Similar heterostructures have previously been rendered for topological insulator superlattices [45] and grapheme nanoribbons [43, 44].

For CDPs to occur, we require two more dimensions added to the wave vector dimension provided by the existing 1-D superlattice. It's noticeable that the coupling of nearest-neighbor TIMs, inclusive of both the magnitude and the sign, can be feasibly tuned by altering the repeated number of the PC's unit cell between adjacent interfaces. Moreover, the on-site resonance frequency of a TIM can be highly controllable if we put a defective unit with adjustable thickness at the interface. Therefore, the modulation of the coupling between adjacent TIMs and the on-site frequency of a single TIM is readily available, which allows us to parameterize these two variables and treats them as two artificial momentum dimensions. Through meticulous design, WPs can thus arise in such 3-D synthetic space owing to the hybrid modes designated as the super-modes, but the realization of CDPs begs for the overlap of two equivalently charged WPs. To this end, we exploit the polarization degree of light. The fact that TIMs response discriminately to TM and TE polarized light appends a so-called 'pseudospin' degree of freedom to the synthetic space, and the appearance of a CDP is finally achieved by merging a pair of WPs with the same topological charge but different pseudospins. Surprisingly, the CDP can conversely be split into two spawned WPs in synthetic space, whose trajectories are tunable via utilizing the pseudospin degree. Such procedure has never been revealed in practice prior to us, offering the evidence that our proposed artificial systems are used to not only explore topological excitations discovered before, but also navigate a way of studying novel phenomena. In particular, we design an applicable and smart strategy to detect the CDP and spawned WPs straightforwardly, which has never been reported before us.

## Section III. Modulations of the TIMs

In this section, we put forward the methods to manipulate the on-site resonance frequencies, coupling effects, and their pseudospin dependence of the TIMs, all of which are essential for the emergence of CDPs as we argued above.

### A. The adjustability of the single TIM

We first provide a detailed introduction to the proposed structure for a single TIM. As shown in Figure 1(a), it consists of PC-p, PC-q, and a defective unit D. The unit cells of PC-p, PC-q, and the defective unit D are represented as $p = M1_{d_1/2} M2_{d_2} M1_{d_1/2}$, $q = M2_{d_2/2} M1_{d_1} M2_{d_2/2}$, and $D = M1_d M2_d$. The stacking structure built up of repeated p(q)-type unit cells can thus be described as $p_6$ ($q_6$), in which the subscripts are employed to show the number of unit cells. Hence, we adopt $p_6 D q_6$ to label the structure shown in the bottom of Figure 1(a), which can support a single TIM [50]. Experimentally, such structure is fabricated with e-beam evaporation and we fabricate three samples with identical $d_1 = 70$nm and $d_2 = 79$nm, but setting $d = 0$nm (sample I), $d = 5$nm (sample II), and $d = 10$nm (sample III), respectively. Figure 1(b) exhibits the scanning electron microscope (SEM) picture of sample II with the highlighted p-type and q-type unit cells. The measured transmission spectra of the sample II (III) are given in Figure 1(c) by black (magenta) circles under normal incident light ($k_x = 0 \mu m^{-1}$), where the common band gaps (bands) of the two PCs are highlighted as the white (grey) regions. It can be seen that sharp peaks inside the gap appear, which are attributed to the excitation of a TIM. To verify this, Figure 1(d) exhibits the calculated spatial distribution of the electric field profile for the associated state of Sample II, from which we can see that such state decays rapidly away from the position of D—a distinctive signature of the TIM. We can also see that the TIM peak of sample III lies at the lower frequency than that of sample II in Figure 1(c), showing the $d$-dependent feature of resonance frequency of TIMs. To make it clear, resonance frequencies of TIMs for these three samples are marked by red open circles in Figure 1(e), decreasing significantly with increasing value of $d$. This confirms the fact that modulating thickness of the defective unit provides us a feasible strategy to adjust on-site frequencies of TIMs.

Furthermore, the TIM exists for both TM and TE polarizations, but they are degenerate in the case of normal incidence ($k_x = 0\mu m^{-1}$). To lift such degeneracy, we need to use oblique incident light with $k_x \neq 0\mu m^{-1}$, and hence we measure the transmission spectra of sample II under the TM (TE) polarized light with oblique incident angles ($k_x = 6\mu m^{-1}$), as shown by red (TM) and blue (TE) open circles in Figure 1(c), confirming the removal of the degeneracy. Figure 1(f) shows the measured TIM frequencies as a function of $k_x$ for Sample II for both polarizations. It is noticeable that the splitting between TM (red) and TE (blue) polarized TIMs increases monotonically with the increment of $k_x$, matching well with the calculated results (solid lines). Therefore, such splitting between TM and TE polarized TIMs affords us another degree of freedom to manipulate the TIMs.

**B. Coupling signs and magnitudes between two TIMs**

Next, we investigate the effects of q-type and p-type PCs as the coupling channel between two TIMs. As shown schematically in Figure 2(a), each structure is made up of stacking PCs (p-type and q-type) separated by two defective units. With the same notation in Figure 1(a), two designs in Figure 2(a) can then be denoted as $p_6Dq_NDp_6$ and $q_6Dp_NDp_6$, respectively. The overlapping of two individual TIMs with the same frequency $\omega_0$ gives rise to two hybridized TIMs, one symmetric mode (S) at $\omega_S$ and one anti-symmetric one (AS) at $\omega_{AS}$. Here, the symmetric types are defined by the symmetry of the electric field which uses the center of $p_N$ ($q_N$) as the reference point. In the following, we demonstrate that the normalized coupling strength $J \equiv (\omega_S - \omega_{AS})/2\omega_0$, which describes the coupling amplitudes and signs, is directly controlled by $N$ for either q-type or p-type PC in the middle.

We start by considering the q-type PC as the coupling elements, namely $p_6 D q_N D p_6$ with normal incidence ($k_x = 0 \mu m^{-1}$). Figure 2(b) shows the transmission spectra of $p_6 D q_6 D p_6$ and $p_6 D q_7 D p_6$ (black circles). For both samples, we see two transmission peaks owing to two hybridized TIMs (S and AS). For the sample $p_6 D q_6 D p_6$, $\omega_S < \omega_{AS}$ such that $J < 0$. While for the sample $p_6 D q_7 D p_6$, $\omega_S > \omega_{AS}$ such that $J > 0$. In Figure 2(c), the $N$-dependence of $J$ for q-type PC case, which is extracted from experimental data, is shown by blue downward-pointing triangles. We can see that $|J|$ possesses a negative association with $N$, and the sign of $J$ totally relies on the parity of $N$. For the samples $p_6 D q_N D p_6$, if $N$ is odd, $J < 0$, otherwise $J > 0$. This is because the accumulated phase for each unit cell is $\pi$.

We then explore $J$ for the p-type PC as the coupling channel case, namely the samples $q_6 D p_N D p_6$. The corresponding results are present as red upward-pointing triangles in Figure 2(c), in which $J$ has the same magnitude as that of $p_6 D q_N D p_6$ but with opposite signs for a given $N$. In order to confirm these results, we also plot the results calculated by COMSOL in Figure 2(c), which shows good agreements with experimental ones. The sign of $J$ is determined by the coupling between two TIMs. More details of two coupled TIMs are provided in the section I of the Supplementary Materials.

Moreover, we investigate coupling effects for nonzero in-plane wave vector ($k_x > 0 \mu m^{-1}$). Figure 2(b) also depicts the measured transmission spectra of the two samples $p_6 D q_6 D p_6$ and $p_6 D q_7 D p_6$ for TM (red open circles) and TE (blue open circles) polarizations with $k_x = 6 \mu m^{-1}$, indicating the polarization-dependent characteristic of the hybridized TIMs. To get a further step, we plot $J$ as a function of $k_x$ in Figure 2(d), where $p_6$ ($q_6$) and $p_7$ ($q_7$) are employed as representatives of the even and odd $N$ cases. The result clearly shows that the sign of $J$ changes in the same way as that of $k_x = 0 \mu m^{-1}$. However, given a fixed $p_N$ ($q_N$), the magnitude of $J$ due to the TM mode has discriminated variation tendency compared with $J$ of the TE one as $k_x$ increases, which almost remains the same for the former while decreases significantly for the latter. The calculated results by COMSOL are shown in Figure 2(d) by lines, which used to confirm the experimental results.

**Section IV. Realization of the CDPs and spawned WPs in synthetic space.**

According to the previous analysis, the eigenfrequency of a single TIM is readily controllable, and the coupling (including signs and magnitudes) between two adjacent TIMs is highly tunable. All of these are sufficient for us to construct a 1-D topological superlattice analogous to a dimerized atomic chain, where we regard TIMs as photonic orbitals. The hybridization of them forms hybrid orbitals, which are referred to as super-modes here. As a consequence, we deliberately design an optical superlattice to create a periodic sequence of TIMs, which is built up of alternating structures of $p_m$, $D_A$, $q_n$, and $D_B$, as illustrated in Figure 3(a). The *i*-th unit dimer with two sublattice sites $A_i$ and $B_i$ is defined as $[p_{m/2}D_A q_n D_B p_{m/2}]$ marked by the magenta dashed rectangle in Figure 3(a). In this notation, the subscripts A and B denote two different defective units with their respective thickness $d_A$ and $d_B$, and the subscripts $m(n)$ labels the number of unit cells of the q(p)-type PC. We then express the on-site resonance frequencies of two adjacent TIMs as $\omega_{A,s} = \omega_s + \Delta_s$ and $\omega_{B,s} = \omega_s - \Delta_s$, where $s=\uparrow\downarrow$ denoting two polarizations, $\omega_s \equiv (\omega_{A,s} + \omega_{B,s})/2$, and $\Delta_s \equiv (\omega_{A,s} - \omega_{B,s})/2$. Here, $\Delta_s$ refers to a staggered on-site frequency offset regarding $\omega_s$ as a reference value. As shown in Figure 1, the values of $\Delta_s$ are determined by the difference $d_A - d_B$. The coupling PCs $q_n$ ($p_m$) directly determine the intra (inter)-dimer coupling strength, denoted as $J_{1,s}$ ($J_{2,s}$). What's more, a remarkable feature of our superlattice system is the adjustability of the coupling sign, since that $m$ and $n$ are simultaneously odd or even numbers leads to $J_{1,s}J_{2,s} < 0$, otherwise $J_{1,s}J_{2,s} > 0$. Taking this into account, we introduce an additional parameter $g$ as $g \equiv \text{sgn}\left(\dfrac{J_{2,s}}{J_{1,s}}\right)$, and utilize $J_s \equiv (-gJ_{1,s} - J_{2,s})/2g$ and $\delta_s \equiv (-gJ_{1,s} + J_{2,s})/2g$ for further discussion. As shown in Figure 2, the values of $\delta_s$ are determined by $m-n$. Accordingly, the Hamiltonian for the super-modes formed by multiple TIMs can be written as an effective dimerized model

$$H = \sum_{i,s=\uparrow,\downarrow} -(J_s+\delta_s)a^\dagger_{i,s}b_{i,s} - g(J_s-\delta_s)a^\dagger_{i+1,s}b_{i,s} + h.c. \\ +(\omega_s+\Delta_s)a^\dagger_{i,s}a_{i,s} + (\omega_s-\Delta_s)b^\dagger_{i,s}b_{i,s} \quad (1)$$

Here, $a^\dagger_{i,s}$ ($b^\dagger_{i,s}$) and $a_{i,s}$ ($b_{i,s}$) are the creation and annihilation operators of the TIM lying on $A_i$ ($B_i$) site of the chain, respectively. Since $D_A$ and $D_B$ have negligible effect on the coupling strength, $\delta_s$ and $\Delta_s$ can be treated as independent parameters. If we merely restrict ourselves to the case of zero in-plane wave vector ($k_x = 0\mu m^{-1}$), this Hamiltonian represents a novel 1-D Su-Schrieffer-Heeger (Rice-Mele) chain with $d_A = d_B$ ($d_A \neq d_B$) and hence $\Delta_s = 0$ ($\Delta_s \neq 0$), of which band structures and topological properties such as topological end states are analyzed detailedly in the section II of the Supplementary Materials. Based on the fact that the degeneracy of TM and TE polarized TIMs lifts when $k_x \neq 0\mu m^{-1}$, $\omega_s$ is a function of $k_x$ such that $\omega_s(k_x) = \omega_0 + \tau_s(k_x)$, where $\omega_0 \equiv \omega_s(0)$ denotes the eigenfrequency of the $k_x = 0\mu m^{-1}$ case and $\tau_s(k_x)$ refers to the frequency shift compared with $\omega_0$. The Hamiltonian (1) can thus be transformed into the Bloch momentum space, and expressed in the pseudospin up ($s=\uparrow$) (TM) and down ($s=\downarrow$) (TE) representation as

$$H = \tilde{\tau}\sigma_z \otimes \sigma_0 + \begin{pmatrix} \tilde{d}_\uparrow \cdot \sigma & 0 \\ 0 & \tilde{d}_\downarrow \cdot \sigma \end{pmatrix}. \quad (2)$$

Here, we introduce $(\tilde{d}_s)_x = -(J_s+\delta_s) - g(J_s-\delta_s)\cos\xi\Lambda$, $(\tilde{d}_s)_y = -g(J_s-\delta_s)\sin\xi\Lambda$, $(\tilde{d}_s)_z = \Delta_s$, and $\tilde{\tau} = \frac{1}{2}(\tau_\uparrow(k_x) - \tau_\downarrow(k_x))$, in which $\Lambda$ is the length of the unit dimer, $\xi$ serves as the Bloch wave vector in the $z$ direction and $\sigma$ stands for Pauli matrices. Thereby, the eigenvalue of the Hamiltonian (2) could be figured out, denoted as $\tilde{\omega} \equiv \omega - \overline{\omega_0}$ with $\overline{\omega_0} = \omega_0 + \frac{1}{2}(\tau_\uparrow(k_x) + \tau_\downarrow(k_x))$.

With respect to the special case at $k_x = 0\mu m^{-1}$, TM and TE polarized super-modes are degenerate since $\tilde{\tau}=0$ and $\tilde{d}_\uparrow = \tilde{d}_\downarrow$. Hence we introduce the parameters $\delta \equiv \delta_\uparrow = \delta_\downarrow$ ($k_x = 0\mu m^{-1}$) and $\Delta \equiv \Delta_\uparrow = \Delta_\downarrow$ ($k_x = 0\mu m^{-1}$), together with the original Bloch wave vector $\xi$, to form a 3-D synthetic space $(\delta, \xi, \Delta)$. The Hamiltonian then can be transformed into $H(\delta, \xi, \Delta) = (H_\uparrow, 0; 0, H_\downarrow)$, in which $H_{s=\uparrow,\downarrow} \equiv \tilde{d}_{s=\uparrow,\downarrow} \cdot \sigma$. As a result, the associated four bands cross at the degenerate point $(\delta, \xi, \Delta) = (0,0,0)$. To characterize this degenerate point, we expand the two-by-two Hamiltonian $H_s$ around it:

$$H_s = \delta v_{\delta x,s} \sigma_x + \xi v_{\xi y,s} \sigma_y + \Delta v_{\Delta z,s} \sigma_z, \tag{3}$$

where $v_{\delta x,s} = -2$, $v_{\xi y,s} = \Lambda J_s$, and $v_{\Delta z,s} = 1$ (See details in the section III of the Supplementary Materials). The above Hamiltonian exhibits a standard Weyl Hamiltonian form, and thus the band crossing point for either TM or TE super-modes can be regarded as a WP in the synthetic space. An important characteristic of a WP is the capacity to carry a topological charge, which corresponds to its chirality $c_s$ ($=\pm 1$). The Hamiltonian (3) possesses the form of $H(q) = q_i v_{ij} \sigma_j$ with $c_s = \text{sgn}(\det[v_{ij,s}])$, indicating that $c_s$ is equal to $-\text{sgn}(J_s)$. This shows that the chirality $c_s$ of WPs relies on the sign of $J_s$, which is decided by the parity of *m(n)* (See in the section III of the Supplementary Materials). According to the degeneracy of TM and TE polarized super-modes when $k_x = 0\mu m^{-1}$, $c_\uparrow = c_\downarrow$ such that the four-band Hamiltonian indicates an overlapping of two WPs with the same topological charge in synthetic space. Shown in Figure 3(c) as a transparent blue cone in the $\delta - \Delta$ space at $\xi = 0$, such kind of band crossing is known as Charge-2 Dirac point (CDP), whose Hamiltonian is the direct sum of two identical spin-1/2 WPs at the Brillouin zone center and thus has a Chern number of $\pm 2$, contrary to a conventional 3-D Dirac point consisting of two WPs with opposite Chern numbers. The band dispersion in the $(\xi, \Delta) = (0,0)$ plane (highlighted by black solid lines in Figure 3(c) is illustrated in Figure 3(e), showing linear property adjacent to the degenerate point. Through such a way, we have provided a novel method to realize the generalized CDPs with $c = \sum_s c_s = \pm 2$ in the optical frequency regime by manipulations of 1-D optical superlattices exploring the concept of synthetic dimensions.

Such a four-fold cone can be detected unambiguously in experiment. We start by making five samples with structural parameters $(m,n)=(4,4),(4,6),(4,8),(6,4)$, and $(8,4)$, respectively, featured by $d_A = d_B$ such that $\Delta = 0$. We then measure the transmission spectra under normal incidence for these five samples to obtain locations of $\tilde{\omega}$. Figure 3(b) presents the transmission spectra as a function of $\Delta f$ (utilizing $\overline{\omega_0}$ as the reference) for the sample $(m,n)=(6,4)$, where the black dashed lines emphasize the super-modes band edges. In Figures 3(c) and 3(e), we employ black squares to mark locations of such band edges, which almost lie on the crossing lines indicating a great agreement with the theory. The locations of $\tilde{\omega}$ for other four samples are plotted as well by different dots in Figures 3(c) and 3(e), all of which are situated at the crossing lines and thus exhibit the characteristic of linear crossing for the Dirac point, matching with the theory quite well (See details in the section IV of the Supplementary Materials). Moreover, we fabricate another two samples with $(m,n)=(4,6)$ and $(6,4)$, characterized by $d_A \neq d_B$ and hence $\Delta \neq 0$. The results gotten from experimental data are also shown in Figures 3(c), well-located at the cone's surface. Consequently, the experimental results support our theory of the CDP, and hence the realization of a CDP in the visible light range is achieved.

When $k_x \neq 0\mu m^{-1}$, the degeneracy of TM and TE super-modes is removed, lending to $\tilde{\tau} \neq 0$ and $\tilde{d}_\uparrow \neq \tilde{d}_\downarrow$. Therefore, nonzero $k_x$ will split the CDP at $k_x = 0\mu m^{-1}$ into two WPs of TM and TE polarized super-modes, respectively. The solid surface in Figure 3(d) shows such two WPs in the $\delta - \Delta$ space at $\xi=0$ with $k_x=6\mu m^{-1}$, and the dispersion in the $(\xi,\Delta)=(0,0)$ plane are present in Figure 3(f) by red(blue) solid lines for TM(TE) modes. To demonstrate it, we measure the transmission spectra under oblique incident light of the seven samples defined in Figures 3(c) and 3(e). We choose to show the transmission spectra of the sample $(m,n)=(6,4)$ in Figure 3(b), and the band edges of the super-modes for TM and TE polarizations are highlighted by red and blue vertical lines. We see that the band edges red-shifted (blue-shifted) for TE (TM) polarization, which agrees with theoretical results in Figure 3(d). We further mark the locations of associated $\tilde{\omega}$ of all these samples in Figures 3(d) and 3(f) with red (blue) color for the TM (TE) super-modes. The consistency between the theory and experiments indubitably corroborates our idea that the CDP is separated into two WPs in the synthetic space with the frequency splitting equals $2\tilde{\tau}$ resulting from the nonzero $k_x$. As a result, $\tilde{\tau}$ can be treated as the effective Zeeman term, which increases with the enhancement of the "magnetic field" (that is, the increase of $k_x$). Note that varying $k_x$ has no effect on $c_s$ of both WPs, so $c_\uparrow = c_\downarrow$ as those of $k_x = 0\mu m^{-1}$ (See details in the section III of the Supplementary Materials).

**Section V. Topological end modes in truncated optical superlattices.**

In contrast to the conventional 3-D DPs, which carry no net topological charge and thus are lack of topological surface states, the CDPs arising in our system characterized by Chern numbers equal to $\pm 2$ imply the existence of intriguing topological end modes. Such end modes originate from each of the two WPs guaranteed by the bulk-edge correspondence, known as one of the most significant experimental properties of WPs. In our synthetic space, these topologically protected modes can be separated into two groups due to TM and TE polarized super-modes, respectively, each of which is supposed to be engineered independently under different polarizations. In what follows we demonstrate their existence in our optical superlattice system when truncated in space. The configurations could be generalized as $p_s [p_{m/2} D_A q_n D_B p_{m/2}]_5 p_s$, which are composed of 5 unit dimes with an open condition along z-axis and are extended by $s$ additional unit cells of at both termini to avoid interacting with external environment. Figure 4(a) sketches a specific structure with $(m,n,s)=(4,6,5)$ and $d_A \neq d_B$ such that $\delta < 0$ and $\Delta \neq 0$, meeting the condition of supporting two nondegenerate end states (See details in the section V of the Supplementary Materials). The substrate is made from $SiO_2$ at the bottom of the structure. We describe the incident light from the front (bottom) as F (B). Topological end modes should come in pairs regardless of the value of $\Delta$, but it is $\Delta$ that determines locations of these two end states in synthetic space. To make it clear, in Figure 4(d) we depict eigenfrequency surfaces of WPs and corresponding topological ends modes in the δ-Δ space with $k_x=6\mu m^{-1}$ for both TM and TE polarized super-modes. In Figure 4(d), the end modes on purple sheets are located at the front side of truncated chains with $\tilde{\omega}=\pm\tilde{\tau}+\Delta$ and can only be excited by F, whereas those on orange sheets are localized at the end of the chains with $\tilde{\omega}=\pm\tau-\Delta$ excited only by B, in which the first plus (minus) sign applies to TM (TE) polarized end modes. Notably, the end states belonging to the intersections of the two sheets connect to the WPs, sharing the same mathematical origin as that of the Fermi arc surface states [1, 3, 6] in Weyl semimetals. They are plotted by magenta dotted lines, with $\Delta=0$ and $\tilde{\omega}=\pm\tau$ in synthetic space, and hence they can be excited by either F or B. It is, however, indispensable to underscore here that the Fermi arc links two WPs in a periodically arranged system while the Fermi-arc-like end modes in our system connect a WP to the boundary of synthetic space explained by the existence of the net topological charge.

Figure 4(b) presents the measured reflection spectra of the sample illustrated in Figure 4(a) for TM and TE polarizations excited by both F and B with an oblique incident angle of $30°$, where we observe four dips inside the TIM gaps. These modes are labelled as 1 to 4, whose distributions of the electric field norm are exhibited in Figure 4(a). It is worth noting that, for both TM and TE polarized super-modes, the modes excited by F (1 and 3) and B (2 and 4) are located at the front and bottom of the sample, respectively. We then investigate another sample with $(m,n,s)=(4,6,5)$ and $d_A=d_B$, characterized by $\delta<0$ and $\Delta=0$. Its measured reflection spectra achieved from F with an incident angle of $30°$ for TM and TE polarizations are exhibited in Figure 4(c), where each dip in the gap of super-modes is attributed to two degenerate end modes localized at both termini. Figures 4(f) and 4(g) provide projected dispersion cones within different $\Delta$ planes. The eigenfrequency surfaces of topological end modes are projected as straight dashed line and the locations of end modes mentioned above are plotted in corresponding planes. Therefore, the great conformance between calculations and experiments further verifies our argument of topological end modes based on the established optical superlattice system. In addition, the introduction of the nonzero $k_x$ leads to a removal of the degeneracy between TM and TE polarized end modes, each of which divides into two states with the splitting increasing rapidly as $k_x$ increases, as revealed in Figure 4(f) for the sample described in Figure 4(a).

**Section VI. Conclusion & Discussion**

Our work presents a flexible strategy based on elaborately designed 1-D optical superlattices to realize CDPs in 3-D synthetic space. We demonstrate the highly tunable on-site resonance frequency of each TIM and the controlled periodic coupling of nearest-neighbor TIMs within our superlattices, which motivate us to employ them as two parametric dimensions. The TIMs play the role of photonic orbitals, and their hybridizations form topological super-modes, whose band structures can be ingeniously engineered to create CDPs in synthetic space with the pseudospin degree originated from the polarized property of light. It is, for the first time, to realize CDPs in the visible region. Without the help of synthetic dimensions, as well as the utilization of pseudospins which fundamentally change the system's behaviors, the creation of CDPs is more demanding, possible only in the infrared range restrained by obstacles to the fabrication of complex structures. More interestingly, the CDP can be artificially split into two spawned WPs with the same Chern number by introducing nonzero horizontal wave vector that removes the degeneracy between TM and TE polarized super-modes. Such amazing process hasn't been obtained in previous studies, thus opening a new frontier to explore novel emergent phenomena of topological physics. In the past, the detections of topological monopoles are always challenging and need advanced experimental techniques, but the approach we render here is facile and obtainable by measuring transmission spectra to examine band structures of super-modes unambiguously. In addition, the bulk-edge response due to the CDP displays itself as topological end modes residing at boundaries of truncated superlattices, which can be manipulated with ease and hence be applied for local field enhancement in various realms [47-49].

An impressive result achieved in our work is that the sign of the coupling can be either positive or negative. There are a variety of exotic phenomena emerging due to the mixture of positive and negative couplings between physical states [51-53], which are always costly to realize in real space. The realization of such couplings in terms of our system is incredibly convenient and straightforward, showing a great advantage in applications. The geometric parameters $m$, $n$, $d_A$ and $d_B$ are specific in this work, as well as the refractive indexes of dielectric materials. However, all of these parameters can be flexibly controlled via existing experimental techniques [54, 55], offering us the ability to continuously tune the coupling and the on-site frequency, namely, $\delta$ and $\Delta$ in synthetic space. Moreover, the topological notions can be extended to all frequency bands, and thus we are likely to study high-frequency physics of topology utilizing similar strategies.

**Acknowledgements**

H. L. thanks C. T. Chan for helpful discussions. H. L. gratefully acknowledges the support of the National Key Projects for Basic Researches of China (Grants No. 2017YFA0205700 and No. 2017YFA0303700), and the National Natural Science Foundation of China (Grants No. 11690033, No. 61425018, No. 11621091, and No. 11374151). K. D. acknowledges funding from the Gordon and Betty Moore Foundation.

M. H. proposed and designed the system. M. H., T. Q., and X. J. carried out the experiments. M. H., K. D., Q. W., and H. L. contributed to the experimental characterization and interpretation and developed the theory. M. H. and K. D. co-wrote the manuscript. All of the authors were involved in the discussions. M. H. and K. D. contributed equally to this work.

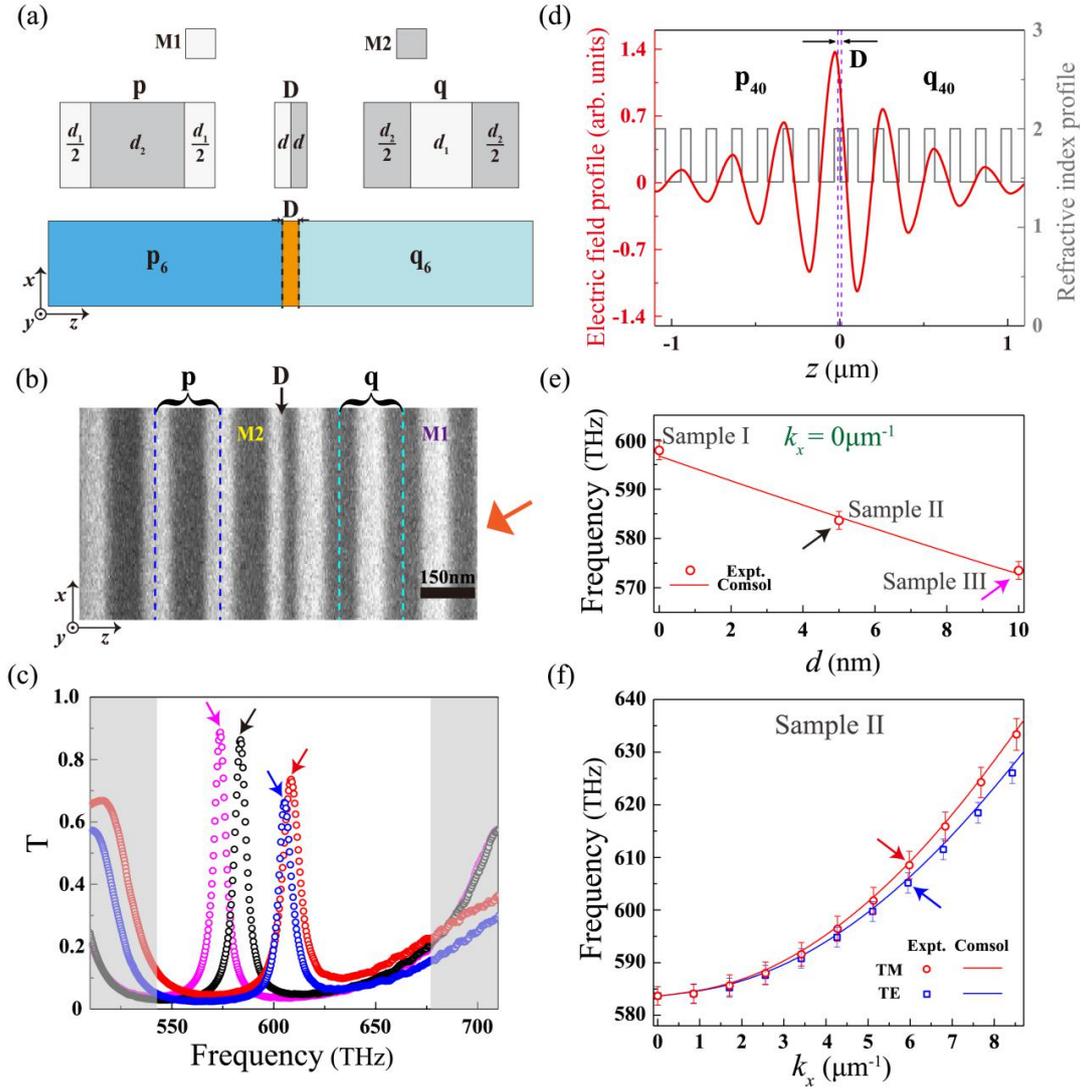

**FIG. 1.** (a) The proposed structure for a single TIM. The upper panel shows the unit cell of both PCs (p-type and q-type unit cell) and the configuration of the defective unit. All of them are made up of alternating layers of two dielectric materials, denoted as M1 (white) and M2 (gray). The thickness notations are indicated on each layer. (b) SEM picture of Sample II. The orange arrowhead shows the direction of incident light. Here, we employ $HfO_2$ as M1 (bright region) and $SiO_2$ as M2 (dark region). The refractive index of $HfO_2$ and $SiO_2$ are 2 and 1.46, respectively. (c) Measured transmission spectra of Sample II under normal incidence, oblique incidence of TM waves, and oblique incidence of TE waves are shown by black, red, and blue circles, respectively. The oblique incident angle is 30°. The magenta circles show measured

transmission spectrum of Sample III under normal incidence. (d) The calculated electric field profile of the TIM for Sample II under normal incidence is plotted by the solid red line. The grey line shows the corresponding refractive index profile. (e) $d$-dependent resonance frequency of the TIM with $k_x = 0\,\mu\text{m}^{-1}$. (f) The in-plane dispersion relation of the TIM for the TM (red line) and TE (blue line) polarizations excited in sample II. In both (e) and (f), solid lines are calculated by COMSOL, and open markers are obtained directly from experimental data. The corresponding experimental transmission spectra of the black, red, blue, and magenta arrowheads in (e-f) are shown in (c). The uncertainties in the measured data are shown by the error bars.

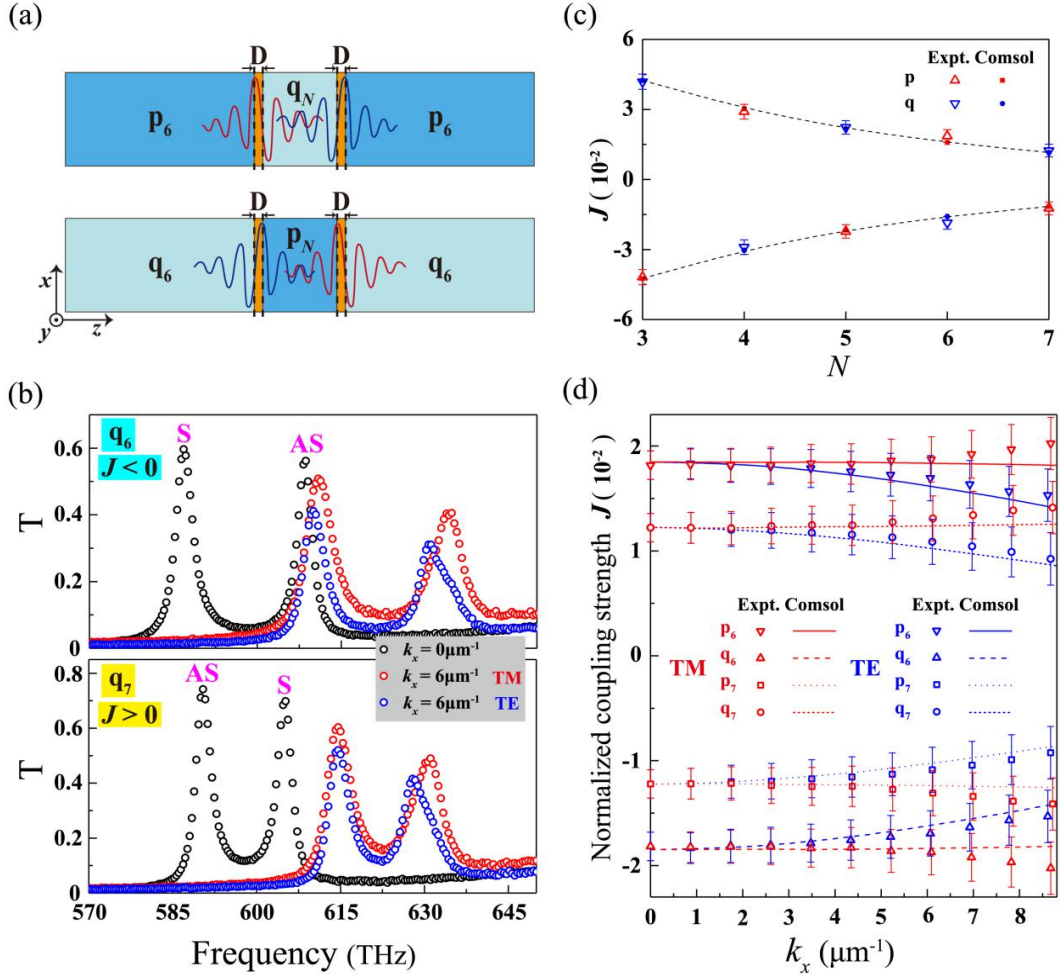

**FIG. 2.** (a) Sketches of two different designs of coupled TIMs (see text). The red and blue curves show schematic mode profiles of two individual TIMs. (b) Measured transmission spectra of $p_6Dq_6Dp_6$ (upper panel) and $p_6Dq_7Dp_6$ (lower panel) with $k_x = 0\mu m^{-1}$ and $k_x = 6\mu m^{-1}$ (incident angle $30°$) for TM/TE polarizations. The parameters of both p-type and q-type unit cells are identical to those in Fig. 1(b) except that $d$ of the defective unit is 0nm. (c) Normalized coupling strength $J$ as a function of $N$ for both p-type (red markers) and q-type (blue markers) PCs. The filled squares (p-type) and circles (q-type) are calculated by COMSOL, whereas open upward-pointing (p-type) and downward-pointing (q-type) triangles are extracted from experimental data. The downward tendency of $|J|$ versus $N$ is shown by dotted lines. (d) Dependence of the normalized coupling strength $J$ on the in-plane wave

vector $k_x$ for both TM and TE polarizations. All lines are calculated by COMSOL, while open markers are obtained from experimental data (see text). Also shown are experimental uncertainties in (c) and (d) by error bars.

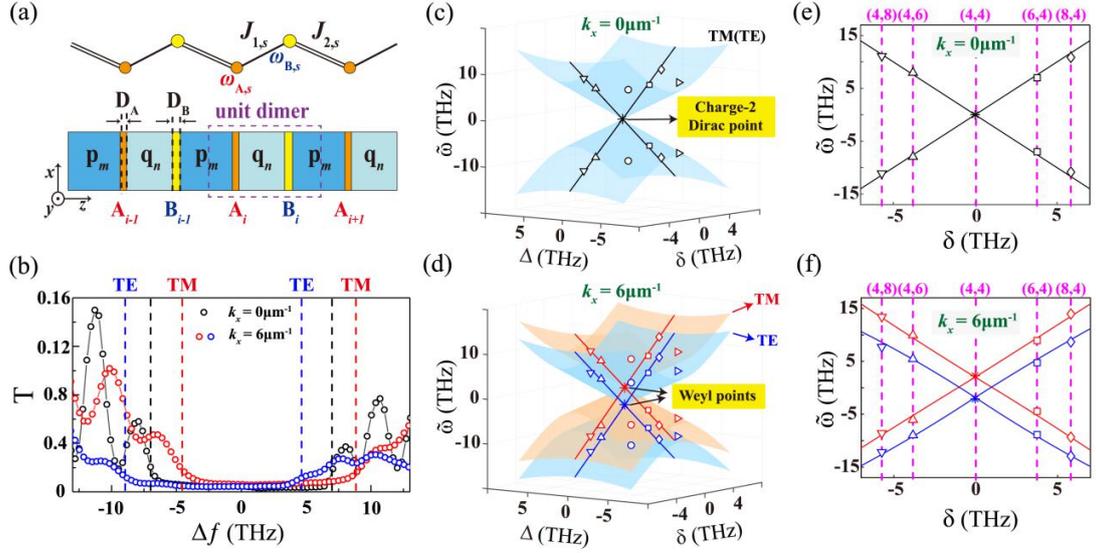

**FIG. 3.** (a) The superlattice of TIMs is shown in the lower panel, and the schematic representation of effective dimerized model is given in the upper panel. (b) Measured transmission spectra as a function of $\Delta f$ for the sample with $(m,n)=(6,4)$ and $(d_A, d_B)=(0,0)$ ( $(\delta,\xi,\Delta)=(3.8,0,0)$ ) under normal incidence ( $k_x = 0\mu m^{-1}$ ) and oblique incidence of TE/TM waves ( $k_x = 6\mu m^{-1}$ ). (c-d) Eigenfrequency surface in the $\delta$-$\Delta$ space at (c) $k_x = 0\mu m^{-1}$ and (d) $k_x = 6\mu m^{-1}$. The surfaces are calculated by the effective dimerized model. The upward-pointing triangles, downward-pointing triangles, squares, and rhombuses mark the bulk band edge frequencies obtained from experimental data of five samples with $(m,n)=(4,6)$, $(4,8)$, $(6,4)$, and $(8,4)$, respectively. The asterisks stand for the center of two gapless bands of super-modes by measuring transmission spectra of the sample with $(m,n)=(4,4)$. The thickness of defective units (expressed in nanometers) for all the five samples are $(d_A, d_B)=(0,0)$. Their associated synthetic coordinates $(\delta,\xi,\Delta)$ endowed with units of $(THz, \mu m^{-1}, THz)$ are equal to $(0,0,0)$, $(-3.8,0,0)$, $(-5.7,0,0)$, $(3.8,0,0)$, and $(5.8,0,0)$. The open circles and right-pointing triangles also represent experimental band edge frequencies of two samples with $(m,n)=(4,6)$ and $(6,4)$,

and the thickness of their defective units are $(d_A, d_B) = (3, 0)$, corresponding to $(\delta, \xi, \Delta) = (-3.8, 0, -4.1)$ and $(3.8, 0, -4.1)$, respectively. Solid lines highlight the $\Delta = 0$ plane. (e-f) Eigenfrequency as a function of $\delta$ at $\xi = 0$ and $\Delta = 0$ when (e) $k_x = 0\mu m^{-1}$ and (f) $k_x = 6\mu m^{-1}$. Solid lines are calculated by the effective dimerized model, and open markers are experimental results with their samples labelled on the top. The unit cells of p-type and q-type PCs for all of samples are described in the section IV of the Supplementary Materials.

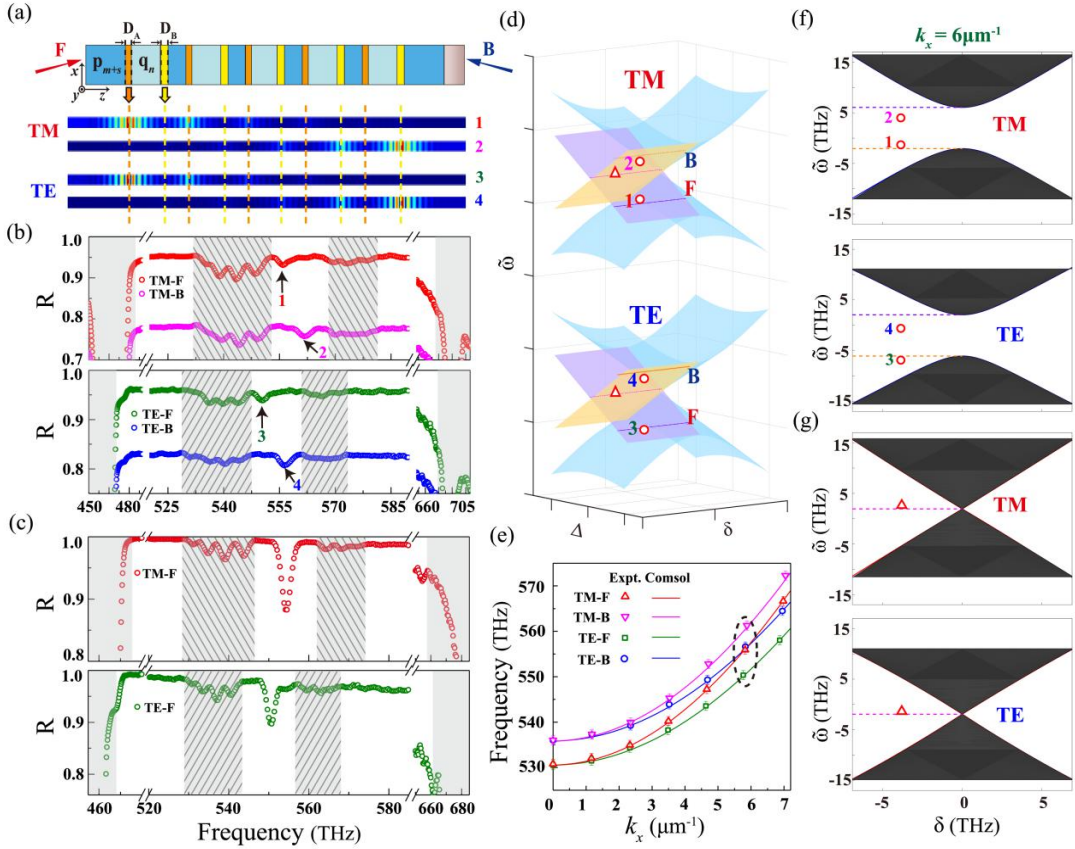

**FIG. 4.** (a) The upper panel: the schematic picture of the truncated optical superlattice $p_5[p_2D_Aq_6D_Bp_2]_5p_5$ with $(d_A, d_B) = (3, 0)$ and $(\delta, \xi, \Delta) = (-3.8, 0, -4.1)$. The lower panel: distributions of the electric field norm of topological end mdoes 1 to 4. (b) Measured reflection spectra of the sample depicted in (a) for TM (upper panel) and TE (lower panel) polarizations excited by both F and B oblique incident light with an angle of $30°$. The four dips marked with black arrows correspond to topological end states 1 to 4. (c) Measured reflection spectra of the sample $p_5[p_2D_Aq_6D_Bp_2]_5p_5$ with $(d_A, d_B) = (0, 0)$ ( $(\delta, \xi, \Delta) = (-3.8, 0, 0)$ ) for TM (upper panel) and TE (lower panel) polarizations excited only by F incident light with an angle of $30°$. In (b-c), transparent gray regions correspond to the common bulk band gaps of PC-p and PC-q, and the gray regions with extra inclined downward and upward lines stands for the bands of TM and TE polarized super-modes, respectively. (d) Eigenfrequency surfaces showing two WPs (transparent blue cones) and topological ends modes

(purple and orange sheets) in the $\delta$-$\Delta$ space with $k_x=6\mu m^{-1}$ for both TM and TE polarizations. The intersections of the two sheets accompanied by each WP are perceived as Fermi-arc-like surface states. For display purpose, the vertical distance between two WPs is deliberately magnified. (e) Frequencies of topological end modes for TM and TE polarizations as a function of $k_x$ for the sample used in (a-b). The solid lines are results of numerical calculations, the open markers are obtained directly from experimental data, and experimental uncertainties are shown by error bars. The topological end states 1-4 shown in (b) are encircled by a black dashed ellipse. (f-g) Eigenfrequencies of topological end states for TM (upper panel) and TE (lower panel) polarizations in the (f) $\Delta=-4.1\text{THz}$ and (g) $\Delta=0\text{THz}$ plane. In (d,f,g), the locations of mode 1-4 in (a-b) and the degenerate states in (c) are marked by circles and upward-pointing triangles, respectively. The black regions in (f-g) refer to the bands of super-modes and the dashed lines in (d,f,g) correspond to the calculated dispersion of topological end modes.